\documentclass[
 reprint,amsmath,amssymb,aps,prl,
 superscriptaddress,
]{revtex4-2}

\usepackage{graphicx}
\usepackage{dcolumn}
\usepackage{bm}
\usepackage{hyperref}

\begin{document}

\preprint{APS/123-QED}

\title{Unconventional superconductivity from lattice quantum disorder}

\author{Yu-Cheng Zhu}
\author{Jia-Xi Zeng}

 \affiliation{State Key Laboratory for Artificial Microstructure and Mesoscopic Physics, Frontier Science Center for Nano-optoelectronics and School of Physics, Peking University, Beijing 100871, People's Republic of China}
 
\author{Xin-Zheng Li}
 \email{xzli@pku.edu.cn}
\affiliation{State Key Laboratory for Artificial Microstructure and Mesoscopic Physics, Frontier Science Center for Nano-optoelectronics and School of Physics, Peking University, Beijing 100871, People's Republic of China}
\affiliation{Interdisciplinary Institute of Light-Element Quantum Materials, Research Center for Light-Element Advanced Materials, and Collaborative Innovation Center of Quantum Materials, Peking University, Beijing 100871, People's Republic of China}

\begin{abstract}
Unconventional superconductivity presents a defining challenge in physics.
Prevailing theoretical frameworks have predominantly emphasized electrons, largely neglecting the rich physics inherent in the lattice. 
%
%
Conventional phonon theory omits quantum many-body effects of the nuclei, leading to misleading structural phase diagrams and an unsound foundation for superconducting theory.
Here, by incorporating nuclear quantum many-body effects within first-principles calculations, we discover a lattice quantum disordered phase in superconductors $\textup{H}_3\textup{S}$ and $\textup{La}_3\textup{Ni}_2\textup{O}_7$. 
This phase occupies a triangular region in the $P-T$ phase diagram, whose left boundary aligns precisely with $T_c$ of the left flank of the superconducting dome. 
Its $T_c^{\textup{max}}$ coincides with the dome's peak, identifying this phase as both the origin of superconducting transition on the left flank and a key ingredient of the pairing mechanism.
Our findings advance the understanding of unconventional superconductivity and establish lattice quantum disorder as a unifying framework, both for predicting new superconductors and for elucidating phenomena in a broader context of condensed matter physics.
\end{abstract}

\maketitle

The left flank of the superconducting dome—spanning the underdoped or low-pressure regime up to the peak $T_c$—is pivotal for deciphering the mechanism of high-temperature superconductivity\cite{ding_spectroscopic_1996,doi:10.1126/science.1133411,PhysRevLett.110.047004}. 
This region is characterized by an intimate entanglement of magnetic, charge, and structural orders\cite{TomTimusk_1999,RevModPhys.63.239}. 
While such complexity has inspired influential theoretical paradigms, including those based on magnetic fluctuations and strong electron correlations, it has also led to a lack of consensus, obscuring a unified physical picture\cite{doi:10.1126/science.1200181,RevModPhys.84.1383}.
The growing family of unconventional superconductors\cite{Stewart03042017} and advances in first-principles methods\cite{pellegrini_ab_2024} now highlight structural phase transitions as a clarifying lens, given their universal link to competing orders and their essential role in ab initio modeling. 
Nevertheless, prevailing theoretical approaches remain predominantly electron-centric, largely overlooking the rich and decisive physics of the lattice\cite{Setty_2024}.

Recent studies of high-pressure superconductors have opened new avenues for understanding high-temperature superconductivity, bringing structural phase transitions into sharp focus\cite{sun_signatures_2023,Hou_2023,zhu_superconductivity_2024,wang_bulk_2024,li_bulk_2026,PhysRevX.15.021005,drozdov_conventional_2015,duan_pressure-induced_2014}.
In such systems, nuclear quantum effects are known to reshape superconducting phase diagrams by altering structural boundaries\cite{errea_quantum_2016,einaga_crystal_2016}. 
A prominent example is $\textup{H}_3\textup{S}$, where accounting for these effects dramatically lowers the pressure for hydrogen-bond symmetrization relative to classical predictions, shifting the entire superconducting dome into the $\textup{Im}\bar{3}\textup{m}$ phase and challenging the prior two-phase interpretation\cite{errea_quantum_2016,yuan2016,Taureau_2024}. 
While such studies incorporated essential quantum corrections, they did not fully address the many-body nature of the interacting nuclei. 
In our recent work, by rigorously including nuclear quantum many-body effects within a first-principles framework, we have achieved a more accurate determination of the structural phase boundaries and, significantly, revealed a lattice quantum disordered (LQD) phase\cite{zhu_quantum_2025}. 
In this regime, quantum fluctuations stabilize a higher-symmetry disordered state, giving rise to lattice dynamics beyond the conventional phonon picture. 
This finding not only redefines the structural phase diagram but also directly motivates the present investigation into the link between such quantum lattice disorder and the unresolved physics on the left flank of the $T_c$ dome.
The recent discovery of nickel-based superconductors, accompanied by enhanced structural characterization under pressure, provides a timely platform for validating this picture\cite{sun_signatures_2023,Hou_2023,wang_structure_2024,zhu_superconductivity_2024,wang_bulk_2024,li_bulk_2026}.

\begin{figure*}[t]
	\includegraphics[width=18cm]{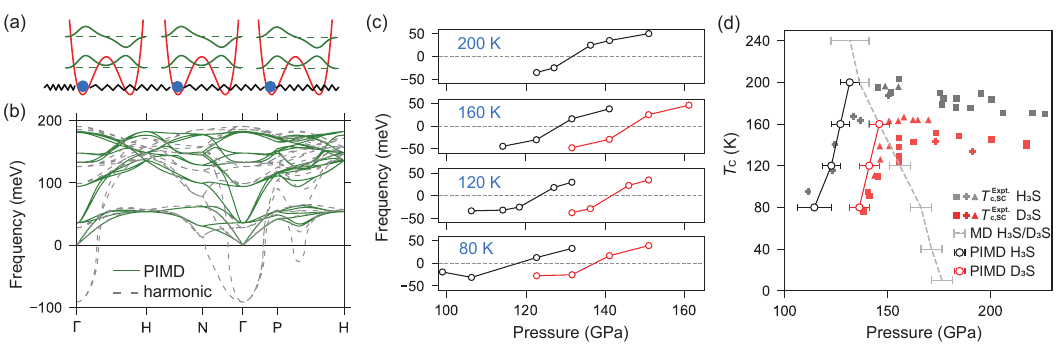}
	\caption{ 
		Lattice quantum disordered phase in $\textup{H}_3\textup{S}$ and $\textup{D}_3\textup{S}$ from first-principles calculation. 
		(a) Schematic of the 1-D double-well chain illustrating the physics of LQD. The nucleus on each lattice site (the ball) lies on a double-well potential (red curve) connected to the neighboring sites by spring interactions (zigzag lines). The many-body nuclear quantum state emerges from the competition between on-site nuclear tunneling and inter-site interactions. In Ref.~\cite{zhu_quantum_2025}, we have a detailed description of this model and the PIMD method.  
		(b) The dispersion relation of lattice dynamics of $\textup{Im}\bar{3}\textup{m}$ $\textup{H}_3\textup{S}$ ($T=200$~K, $P=141$~GPa) by PIMD (solid lines) in comparison with the harmonic phonon spectra (dashed lines). The structural instability indicated by the soft phonon mode is suppressed. 
		(c) PIMD frequency at $\Gamma$ point for the soft mode defined in (b), as a function of temperature and pressure. The open black and red circles correspond to $\textup{H}_3\textup{S}$ and $\textup{D}_3\textup{S}$, respectively. The structural phase transition is defined by the point at which this frequency changes sign. 
		(d) The triangular region of the LQD phase on the $P-T$ phase diagram is bounded by the quantum (PIMD) and classical (MD) phase boundaries. The solid symbols represent the experimental superconducting $T_{c,\textup{SC}}$ from Refs.~\cite{drozdov_conventional_2015,einaga_crystal_2016,D3S2020}. Notably, the left boundary of the LQD phase (the PIMD line) aligns with the left flank of the superconducting dome, and $T_{c,\textup{LQD}}^{\textup{max}}$ coincides with $T_{c,\textup{SC}}^{\textup{max}}$ for both $\textup{H}_3\textup{S}$ and $\textup{D}_3\textup{S}$.
	}
	\label{Fig1}
\end{figure*}

Here, we find that the superconducting transition on the left flank of the dome stems from the structural transition of a low-symmetry phase into a LQD phase. 
We employ path-integral molecular dynamics (PIMD)\cite{chandler1981,marx1996} to treat the nuclear quantum many-body effects from first principles. 
The structural phase boundary is determined by the free-energy surface, which is constructed by the centroid potential of mean force from PIMD. 
The boundary obtained via this approach coincides precisely with the left flank of the superconducting dome. 
Its intersection with the classical boundary from conventional molecular dynamics (MD) delineates the region of the LQD phase. 
The extent of this LQD phase increases linearly with decreasing temperature, with its origin, which we label as $T_{c,\textup{LQD}}^{\textup{max}}$, locating at $\sim$220~K for $\textup{H}_3\textup{S}$, $\sim$160~K for $\textup{D}_3\textup{S}$ and $\sim$77~K for $\textup{La}_3\textup{Ni}_2\textup{O}_7$.
These values align exactly with their respective highest observed superconducting transition temperatures $T_{c,\textup{SC}}^{\textup{max}}$. 
This indicates not only that superconductivity occurs entirely within the high-symmetry phase but also establishes a direct mechanistic link between the LQD phase and unconventional superconductivity. 
We expect this picture to be broadly applicable to other unconventional superconductors.

Thermal and quantum fluctuations are key factors that can suppress a structural phase transition. 
%
A soft phonon mode in the potential energy surface (PES) does not necessarily imply a structural instability in the free energy surface (FES) once these fluctuations are included. 
Thus, the true phase transition point must be identified by examining the curvature of the FES that fully accounts for both thermal and quantum effects. 
This offers a fundamental and broadly applicable criterion for structural instability. 
While thermal fluctuations are well described by conventional MD, the key challenge lies in correctly including nuclear quantum effects\cite{markland_nuclear_2018} when calculating the FES.

At low temperatures, when the potential barrier of a soft mode is modest, nuclear tunneling induces strong quantum fluctuations (sketched by the green wavefunctions of double-well potential in Fig.~\ref{Fig1}a) that favor a quantum disordered state\cite{zhu_quantum_2025}. 
In competition, there is a global tendency for all lattice sites to slide toward the same side of the potential well (as the blue balls) to lower the energy, thereby establishing long-range structural order.
The resulting transition from a low-symmetry to a high-symmetry phase is therefore a quantum order–disorder transition\cite{Sachdev2011}, governed by the balance between quantum fluctuations and long-range ordering (effect of inter-site interaction as the zigzag springs)\cite{zhu_quantum_2025}. 
This physics necessitates a treatment that captures the many-body nature of the quantum nuclei.
We note that ``quantum disorder'' should not be conflated with dynamically or spatially disordered local moments\cite{Taureau_2024}. Outside of model systems like the transverse-field Ising model, this quantum disorder had never been thoroughly and precisely described in real materials, and conventional simulation techniques remained fundamentally unequipped to capture its nuclear quantum many-body nature.

To this end, we have developed a theoretical methodology\cite{zhu_quantum_2025} to construct the FES based on path-integral molecular dynamics (PIMD)\cite{chandler1981,marx1996,markland_nuclear_2018,Cazorla2017}. 
Within PIMD, the centroid potential of mean force provides a rigorous route to construct the relevant FES\cite{cao_formulation_1994,jang_path_1999}. 
While PIMD itself serves as a powerful framework that rigorously incorporates both nuclear quantum many-body effects and thermal effects, the FES approach allows us to correctly extract the critical information regarding structural phase transitions and lattice dynamics that inherently account for these effects.\cite{zhu_quantum_2025}
Leveraging lattice symmetry, the emergence of a soft mode—and hence the phase boundary—can be determined efficiently from the centroid effective forces of only a few finite-displaced structures. 
The region between this quantum phase boundary and the classical boundary from MD defines the LQD phase. 
In the following, we apply this approach to map the LQD phase in two representative systems: $\textup{H}_3\textup{S}$ and $\textup{La}_3\textup{Ni}_2\textup{O}_7$. 
The PIMD simulations were performed using the i-pi software\cite{litman_i-pi_2024}.
To make the PIMD sampling feasible, we employ machine learning inter-atomic potentials at the density-functional theory (DFT) level\cite{behler_generalized_2007,kocer_neural_2022} as the PES, with the underlying DFT calculations performed using the PBE functional\cite{Perdew1996}. 
The computational details are provided in the Supplemental Material\footnote{See Supplemental Material at [URL] for more computational details on PIMD, MD and machine-learning PES, where Refs~\cite{phonopy2023,KRESSE1996,Kresse1999,Zhu2022,PhysRevLett.122.225701,jinnouchi_descriptors_2020,ceriotti_efficient_2010} are also cited.}.

For $\textup{H}_3\textup{S}$, the lattice dispersion obtained via the centroid effective forces (Fig.~\ref{Fig1}b) show that imaginary harmonic frequencies—indicating a saddle point in the PES—are suppressed in the PIMD spectrum when quantum fluctuations are strong. 
Tracking the soft-mode frequency at the $\Gamma$ point across pressure and temperature (Fig.~\ref{Fig1}c) allows us to locate the phase boundary where this frequency changes sign. 
The resulting boundary shifts to higher pressure with increasing temperature (Fig.~\ref{Fig1}d), in contrast to the classical trend and to other quantum correction methods, such as the SSCHA method\cite{errea_quantum_2016,monacelli2021} (see Fig.~S1).
This arises from two competing influences: thermal and quantum effects both suppress structural order, but thermal agitation also disrupts quantum effects, rapidly diminishing the LQD phase. 
Consequently, the structural phase boundary in the $T-P$ plane forms a dome: it first rises with temperature before eventually crossing over to the classical regime. 
The classical boundary, identified from the pair distribution function via MD (Fig.~S2), together with the PIMD quantum boundary, delineates a triangular LQD region. 
Their intersection defines the $T_{c,\textup{LQD}}^{\textup{max}}$ of the LQD phase—the temperature at which quantum effects are completely overwhelmed by thermal fluctuations.
Due to the greater mass of deuterium compared to hydrogen, its quantum effects are weaker. 
As a result, the quantum phase boundary of $\textup{D}_3\textup{S}$ shifts to higher pressures, and its $T_{c,\textup{LQD}}^{\textup{max}}$ is about 160~K, which is lower than the 220~K for $\textup{H}_3\textup{S}$ as in Fig.~\ref{Fig1}d.

In $\textup{H}_3\textup{S}$, the superconducting transition temperature $T_c$ exhibits a characteristic dome in the $P-T$ plane\cite{einaga_crystal_2016}. 
A distinct, continuous kink at the dome’s maximum—evident in the experimental data (the gray filled dots in Fig.~\ref{Fig1}d) and also observed in nickel-based superconductors\cite{sun_signatures_2023}—provides a natural marker that defines the lower-pressure region (below 150 GPa) as the dome’s left flank. 
The key finding is that the left boundary of the LQD phase aligns with the left flank. 
Moreover, the isotope effect in superconductivity is accurately captured by the LQD phase.
Although the quantitative agreement with experiment is limited by DFT precision (results by other functionals are shown in Fig.~S1), 
the correspondence between the phase boundaries strongly indicates that the superconducting transition on the left flank originates from the LQD phase—and, consequently, that 
superconductivity occurs entirely within the high-symmetry $\textup{Im}\bar{3}\textup{m}$ phase, thereby invalidating earlier two-phase interpretations and the calculations based upon them\cite{akashi_first-principles_2015}.

A more important point is that the $T_{c,\textup{LQD}}^{\textup{max}}$ coincides precisely with the peak superconducting temperature $T_{c,\textup{SC}}^{\textup{max}}$ for both $\textup{H}_3\textup{S}$ and $\textup{D}_3\textup{S}$. 
This agreement, robust across different exchange-correlation functionals (Fig.~S1), establishes a direct link between LQD and the superconducting mechanism. 
Having established this framework for $\textup{H}_3\textup{S}$, we now examine $\textup{La}_3\textup{Ni}_2\textup{O}_7$ to test its generality, before finally presenting a unified picture in the subsequent discussion.

\begin{figure}[t]
	\includegraphics[width=9cm]{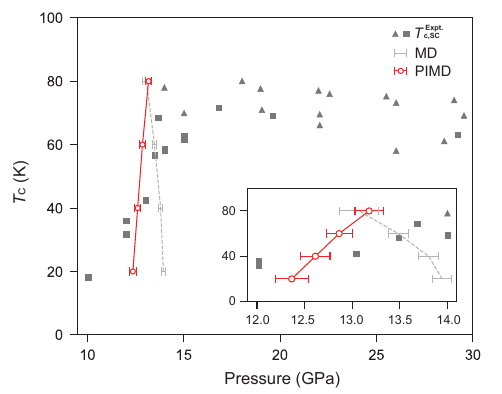}
	\caption{ 
		Lattice quantum disordered phase in $\textup{La}_3\textup{Ni}_2\textup{O}_7$ from first-principles calculation.
		The solid symbols represent the experimental superconducting $T_{c,\textup{SC}}$ from Refs.~\cite{sun_signatures_2023,Hou_2023}. 
		On the left flank of the superconducting dome (below 14 GPa), $T_{c,\textup{SC}}$ is marked by a sudden and rapid decrease. 
		The quantum structural phase boundary by PIMD aligns with the left flank, and $T_{c,\textup{LQD}}^{\textup{max}}\approx77$~K is in consistence with the experimental $T_{c,\textup{SC}}^{\textup{max}}=80$~K.
	}
	\label{Fig2}
\end{figure}

Nickel-based superconductors offer a compelling platform to clarify the link between the LQD phase and unconventional superconductivity. 
In these systems, first-principles calculations have played a central role\cite{sun_signatures_2023}, and the assumed crystal structure—as the starting point of such calculations—directly influences theoretical interpretations of the superconducting mechanism. 
It is therefore critical to correctly determine the structural phase diagram\cite{zhu_superconductivity_2024,wang_bulk_2024,li_bulk_2026}. 
Current studies suggest a transition from I4/mmm to a lower-symmetry phase around a few tens of GPa, and—reminiscent of the earlier, erroneous two‑phase picture for $\textup{H}_3\textup{S}$—assign the left flank of the superconducting dome to that low‑symmetry phase\cite{wang_structure_2024}. 
Here, we resolve the structural transitions in nickel‑based superconductors using $\textup{La}_3\textup{Ni}_2\textup{O}_7$ as a representative example, and reveal the relationship between the LQD phase and the superconducting dome. 
We note that, unlike the earlier studies\cite{sun_signatures_2023,huo_modulation_2025}, we do not employ the PBE+U approach because it fails to reproduce the stable structures and yields phase‑transition pressures far from experimental values (see Fig.~S3). 
For generality and computational feasibility, we use the PBE functional throughout.

As shown in Fig.~\ref{Fig2}, the left boundary of the LQD phase shifts to higher pressure with increasing temperature, matching the steep left flank of the superconducting dome. 
Within the precision of DFT, this indicates that—as in $\textup{H}_3\textup{S}$—the left boundary of the LQD phase coincides with the left flank of the dome, and that superconducting transition on this flank originates from the quantum order‑disorder transition into the LQD phase. 
More importantly, the $T_{c,\textup{LQD}}^{\textup{max}}$ aligns with the $T_{c,\textup{SC}}^{\textup{max}}$ again. 
%
%
The intersection of the PIMD and MD boundaries gives a maximum $T_c$ of about 77~K for the LQD phase of $\textup{La}_3\textup{Ni}_2\textup{O}_7$ at the PBE level, in consistence with the experimental maximum of superconducting $T_c$ of 80~K. 
%

\begin{figure}[t]
	\includegraphics[width=9cm]{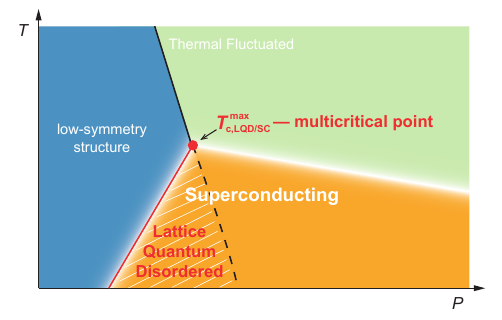}
	\caption{ 
		Schematic phase diagram of lattice quantum disorder and unconventional superconductivity.
		The quantum order-disorder transition boundary (solid red line), dominant at low temperatures, is shifted to lower pressures compared to the classical transition boundary (dashed black line). 
		Their intersection defines $T_{c,\textup{LQD}}^{\textup{max}}$——a multicritical point (red point). 
		The region enclosed by these two boundaries is the LQD phase (shaded orange area). 
		Experimentally observed superconductivity forms a dome (orange area). 
		Crucially, the left flank of the superconducting dome coincides with the left boundary of the LQD phase, and $T_{c,\textup{SC}}^{\textup{max}}$ aligns precisely with the multicritical point ($T_{c,\textup{LQD}}^{\textup{max}}$
		). 
		We contend that this unequivocally points to a superconducting mechanism inherent to the LQD phase, with the multicritical point determining the maximum superconducting transition temperature.
	}
	\label{Fig3}
\end{figure}

Fig.~\ref{Fig3} summarizes the proposed picture for the phase diagram. 
At low temperatures, the structural transition is a quantum order-disorder transition, which shifts the phase boundary to lower pressures compared to the classical limit. 
Above $T_{c,\textup{LQD}}^{\textup{max}}$, the system recovers the classical phase transition behavior. 
Here, we suggest that experiments aiming to locate this boundary should track the change in lattice parameters and target lower temperatures, particularly down to the vicinity of the maximum superconducting $T_c$, rather than at room temperature. 
%
The intersection of the quantum and classical transition lines—which is the $T_{c,\textup{LQD}}^{\textup{max}}$—constitutes a multicritical point. 
Given the observed coincidence between the left boundary of the LQD phase and the left flank of the superconducting dome, we conclude that superconductivity occurs entirely within the high-symmetry phase, thereby refuting earlier two-phase interpretations. 
Crucially, the LQD phase is distinct from an ordered high-symmetry phase (the region on the higher-pressure side of LQD phase); its lattice dynamics, which transcend the conventional phonon picture, may host a novel pairing mechanism.
For instance, as previously discussed by N. M. Plakida $et\ al.$ within the context of the transverse-field Ising model, such lattice motions exhibit a large amplitude and thus induce a strong electron-``phonon'' coupling, thereby enhancing the superconducting $T_c$\cite{Plakida_1987}.

Superconductivity should be viewed not only as a macroscopic quantum state of electronic degrees of freedom, but equally as one of the lattice. 
The energy scale of lattice dynamics (tens to hundreds of kelvins) is comparable to the $T_c$ of high-temperature superconductors; at the very least, the process by which thermal effects suppress nuclear quantum fluctuations cannot be ignored. 
If one attributes the right flank of the dome to a BCS‑like mechanism, where $T_c$ rises slowly with decreasing pressure, the quantum order‑disorder transition could be seen as interrupting this rise. 
However, this cannot explain why the $T_{c,\textup{LQD}}^{\textup{max}}$ coincides precisely with the $T_{c,\textup{SC}}^{\textup{max}}$—that is, why a scenario such as that sketched in Fig.~S4 does not occur. 
Based on consistent first‑principles results across $\textup{H}_3\textup{S}$ and $\textup{La}_3\textup{Ni}_2\textup{O}_7$, we argue that this agreement is not coincidental. 
We instead propose that the right flank of the dome extends from the multicritical point, indicating a profound connection between LQD and the unconventional superconducting mechanism, with the LQD phase playing a decisive role in determining the maximum superconducting transition temperature.

This picture suggests a practical route for predicting and discovering superconductors with higher $T_c$: first identify materials that host a large LQD phase, and subsequently consider appropriate carrier introduction. 
Regarding doping, we expect the LQD framework to remain relevant for doping‑dependent high-$T_c$ families such as the cuprates\cite{keimer_quantum_2015,RevModPhys.75.473}. 
According to the work by Jiang $et\ al.$ investigating the effects of lattice anharmonicity on the superconducting $T_c$ of cuprates\cite{Jiang_2023,Jiang_2024}, it is proposed that doping will influence the soft mode on the potential energy surface in a manner analogous to pressure\cite{Setty_2022}, though its combined effect on both the lattice and carrier concentration is more complex and places higher demands on first‑principles treatments\cite{chen_electronic_2026}. 
Notably, the absence of superconductivity in the antiferromagnetic phase may precisely correspond to the lack of soft modes in the underdoped regime. 
Finally, we suggest that the LQD phase could be a widespread phenomenon, with its extent varying across different materials. 
We anticipate that this perspective will offer fresh insights into other unresolved puzzles in condensed matter, particularly concerning anomalous transport properties such as glass‑like thermal conductivity in crystals\cite{sun2020}.

\begin{acknowledgments}
	We are supported by the National Basic Research Programs of China under Grants No. 2021YFA1400500 and No. 2022YFA1403500, and the National Science Foundation of China under Grants No. 12550005, No. 12404257, No. 12234001, No. 62321004, and No. 12474215.
	The computational resources are provided by the supercomputer center at Peking University, China.
\end{acknowledgments}


\bibliography{ref}

\end{document}